\documentclass[12pt,preprint]{aastex}

\slugcomment{This manuscript is accepted by \textbf{MNRAS}}

\shorttitle{Formation of LMCs in the cores}
\shortauthors{Nejad-Asghar}

\begin{document}

\title{Formation of low-mass condensations in the molecular cloud cores via thermal instability}

\author{Mohsen Nejad-Asghar\altaffilmark{1,2}}

\affil{$^1$Department of Physics, University of Mazandaran,
Babolsar, Iran}

\affil{$^2$Research Institute for Astronomy and Astrophysics of
Maragha, Maragha, Iran}

\email{nejadasghar@umz.ac.ir}

\begin{abstract}
The low-mass condensations (LMCs) have been observed within the
molecular cloud cores. In this research, we investigate the effect
of isobaric thermal instability (TI) applied for forming these LMCs.
For this purpose, at first we investigate the occurrence of TI in
the molecular clouds. Then, for studying the significance of linear
isobaric TI, we use a contracting axisymmetric cylindrical core with
axial magnetic field. Consideration to cooling and heating
mechanisms in the molecular clouds shows that including the heating
due to ambipolar diffusion can lead to the occurrence of TI in a
time-scale smaller than dynamical time-scale. Application of linear
perturbation analysis shows that isobaric TI can take place in outer
region of the molecular cloud cores. Furthermore, the results show
that perturbations with wavelengths greater than few astronomical
units are protected from destabilization property of thermal
conduction, so they can grow to form LMCs. Thus, the results show
that the mechanism of TI can be used to explain the formation of
LMCs as the progenitors of collapsing proto-stellar entities, brown
dwarfs, or proto-planets.
\end{abstract}


\section{Introduction}

The molecular gases have a hierarchical structure which extends from
the scale of the cloud down to much smaller masses of unbound
substructures. The terminology for structure of molecular clouds is
not fixed. The over-dense regions within giant molecular clouds are
termed clumps. The massive clumps are parenting clouds for the
clusters of proto-stellar cores. These dense cores are expected to
undergo gravitational collapse evolving towards singular points and
designated as proto-stars (e.g., Stahler and Palla~2004). Two
important informations intended for the molecular cloud cores are
their shapes and density-profiles. Unfortunately, the core shape
cannot be easily concluded from the apparent observations of
the-plane-of-sky because it is impossible to be directly
de-projected. Although, some works indicate a preference in prolate
cores (e.g., Ryden~1996), but there are also some studies that
favour to a strong preference in oblate cores of finite thickness
(e.g., Tassis~2007). But about the density-profile in the molecular
cloud cores, the observations show the systematically certain
prominent common feature: a close-to-constant density over the
central region ($r < r_{in}$), followed by nearly a power-law
decline over large distances ($r > r_{in}$) (e.g., Hung, Lai and
Yan~2010). According to these observational results, here, we turn
our attention to the molecular cloud cores with cylindrical shape
and appropriate decreasing density-profile.

The observational surveys of the interior of the molecular clouds
have been increased enormously in the last decade, thanks to the
increase of resolution provided by new millimeter and sub-millimeter
interferometers, and also they are due to the systematic combination
of observations of dust and molecular tracers (e.g., Bergin and
Tafalla~2007). In this way, many embedded condensed objects within
each star-forming core have been revealed (e.g., Pirogov and
Zinchenko~2008) which called low-mass condensations (LMCs). They may
be converted to the star-forming gravitationally unstable
proto-stellar cores, brown dwarfs, or proto-planets. Historically,
we can refer to the work of Langer et al.~(1995) who observed LMCs
in the core~D of Taurus Molecular Cloud~1 in the regime of
$0.007-0.021\mathrm{pc}$ and $0.01-0.15 M_\odot$. For the recent
observations, we can refer to the discovery of very low luminosity
objects by the Spitzer 'From Molecular Cores to Planet-Forming
Discs' (c2d) project (e.g. Lee et al.~2009), or the results gained
by Launhardt et al.~(2010) in which they found that at least two
thirds of 32 isolated star-forming cores, which were studied there,
are the evidence of forming multiple stars. Clearly, these
observations can also be used as a witness for existence of LMCs in
the parent core before conversion to the star-forming entities.
Although, due to limitations in resolution and sensitivity of the
available observations (Tafalla~2008), our understanding about
substructures within the cores is still incomplete, but this is well
established that the LMCs are ubiquitous within the molecular cloud
cores.

The LMCs move in the parent core, thus, they may be suitable places
for assembly the dusts and other molecules; they also may be merged
with each other (e.g., Nejad-Asghar~2010). This topic is beyond the
scope of this paper. Here, we want to investigate the theoretical
aspects explaining the formation of LMCs within the cores. Since the
molecular cloud cores are characterized by subsonic levels of
internal turbulence (e.g. Myers~1983, Goodman et al.~1998, Caselli
et al.~2002), small rotational velocity gradients (e.g., Goodman et
al.~1993, Caselli et al.~2002), and subsonic infall motions (e.g.,
Tafalla et al.~1998, Lee, Myers and Tafalla~2001), the turbulence
alone cannot be responsible for the formation of LMCs. Furthermore,
the associated thermal pressure in the cores dominates the turbulent
component by a factor of several (e.g., Tafalla et al.~2004). As
another mechanism for formation of LMCs, we can refer to the work of
van Loo, Falle and Hartquist~(2007) who examined the effect of MHD
waves on dense cores. They found that short-wavelength waves can
play an important role in the generation of LMCs within a core
without breaking it. Here, we can encourage ourselves to
authenticate the suggestion that the thermal instability (TI) may be
a possible favorable scenario for formation of LMCs. This idea can
also be supported by the probes in temperature of the molecular
cores in which the observations show there may be a deviation from
uniform isothermal case (e.g., Harju et al.~2008, Friesen et
al.~2009).

After the pioneer work of Field~(1965), there are many works on
development of TI in the interstellar medium, in both numerical and
analytical approaches. The main idea is that the rapid growth of TI
leads to a strong density imbalance between the cold dense
inhomogeneity and its low-density environment. In the molecular
clouds, Gilden~(1984) calculated the cooling function to show that
molecular gas may be thermally unstable in environment where
$\mathrm{CO}$ cooling dominates. The numerical calculations of
Falle, Ager and Hartquist~(2006) give the idea that TI may have an
important role in formation of internal substructures in the
molecular clouds. Fukue and Kamaya~(2007) considered the effect of
ion-neutral friction of two fluids on the growth of TI. They found
the TI of the weakly ionized plasma in the magnetic field, even at
the small length scales, could be grown up. Nejad-Asghar~(2007)
applied this idea for a one-dimensional self-gravitating magnetized
molecular slab including the frictional heating due to ion-neutral
drift. He found that the heating of ambipolar diffusion in outer
regions of the slab is more significant than the average heating
rates of cosmic rays and turbulent motions, thus, isobaric TI could
take place in this area. Nejad-Asghar and Molteni~(2008) used the
two-fluid smoothed particle hydrodynamics to simulate a partially
ionized one-dimensional cloud. Their gained results indicated that
the TI can insist on the occurrence of density contrast at outer
parts of the molecular cloud cores. Also, the two-dimensional
simulations of Nejad-Asghar and Soltani~(2009) confirmed this idea
that heating of ambipolar diffusion in the molecular cloud cores may
lead to TI and formation of LMCs.

In this paper, we try to make an extension to this idea that TI may
be responsible for the formation of LMCs within the molecular cloud
cores. For this purpose, we consider a cylindrical contracting cloud
as a background unperturbed arrange, with a power-law drop in its
density-profile. In section~2, the net cooling function in the
molecular clouds is calculated, and happening of TI is investigated
too. In section~3, the linear perturbation analysis with Fourier
decomposition of space, is applied on the contracting cylindrical
core to attain the criterion of isobaric TI. The effect of thermal
conduction on TI stabilization is also investigated in Section~3.
Finally, Section~4 is devoted to a summary and conclusions.

\section{The occurrence of TI}

Two key parameters to examine the happening of TI are the net
cooling function and its time-scale. Determination of the cooling
rate for an optically thick, dusty molecular medium is a complex
non-LTE radiative transfer problem. Reviewed article by Dalgarno and
McCray~(1972) presents some of the earlier estimates of cooling rate
by excitation the rotational levels of diatomic molecules. Since at
that time, numerous authors had discussed various aspects of this
problem. For example, Goldsmith and Langer~(1978) analyzed in detail
the cooling produced by line emission from a variety of molecular
and atomic species at temperatures of $10$, $20$, and $40\mathrm{K}$
and for $\mathrm{H_2}$ densities in the range of $10^8 <
n(\mathrm{H_2}) < 10^{12} \mathrm{m^{-3}}$. Hollenbach and
McKee~(1979) evaluated the cooling of most coolant molecules by
using the escape probability approximation which is equivalent to
assumption of constant source function. The more comprehensive study
of radiative cooling rate and emission-line luminosity in dense
molecular clouds carried out by Neufeld, Lepp and Melnick~(1995).
They considered the radiate cooling of fully shielded molecular
astrophysical gas over a wide range of temperatures ($10 < T <
2500~\mathrm{K}$) and $\mathrm{H_2}$ densities ($10^9 <
n(\mathrm{H_2}) < 10^{16} \mathrm{m^{-3}}$). Goldsmith~(2001)
computed the cooling effects of molecular depletion from the gas
phase on grain surfaces in dark clouds. He parameterized the cooling
function as
\begin{equation}\label{cool1}
  \Lambda_{(n,T)} = \Lambda_{(n)} \left ( \frac{T}{10 \mathrm{K}}
  \right ) ^ {\beta_{(n)}},
\end{equation}
and gave the value of parameters to different depletion runs. This
equation was used by Nejad-Asghar~(2007) to investigate the
formation of fluctuations in a one-dimensional molecular cloud
layer.

\begin{figure} \epsscale{0.8} \plotone{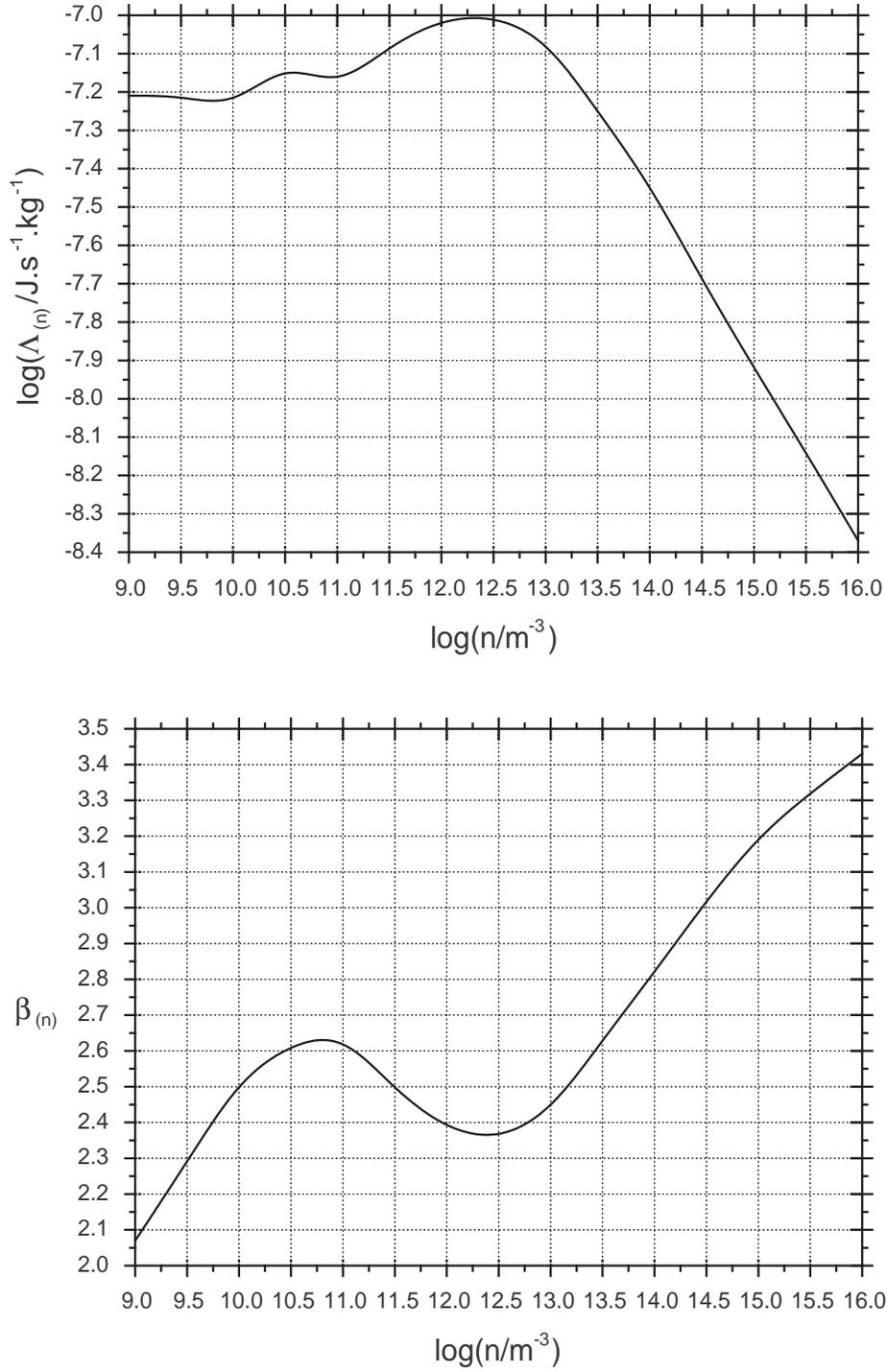}
\caption{The parameters of total cooling rate in the form of
$\Lambda_{(n,T)} = \Lambda_{(n)} (T/10\mathrm{K})^{\beta_{(n)}}$
for molecular clouds with temperatures between
$10-200~\mathrm{K}$.\label{fitlambda}}
\end{figure}

In this paper, we use the cooling function based on the work of
Neufeld, Lepp and Melnick~(1995) which allows us to include cooling
from potentially important coolants of five molecules and two atomic
species: $\mathrm{CO}$, $\mathrm{H_2}$, $\mathrm{H_2O}$,
$\mathrm{O_2}$, $\mathrm{HCl}$, $\mathrm{C}$, and $\mathrm{O}$. The
results presented in figures $3a-3d$ of Neufeld, Lepp and
Melnick~(1995) are convenient to do rough parametrization like the
equation (\ref{cool1}), specially for temperatures between
$10\mathrm{K}$ and $250\mathrm{K}$. The effects of chemistry upon
the cooling rate are dramatically illustrated by the fraction of the
cooling rate attributed to $\mathrm{H_2O}$, which rises rapidly
above temperature $\sim 300~\mathrm{K}$ due to an increase in water
abundance. Thus, we focus our attention to the molecular clouds with
temperature less than $200~\mathrm{K}$. For this temperature range,
the values of parameters $\Lambda_{(n)}$ and $\beta_{(n)}$ are given
in the Fig.~\ref{fitlambda}.

\begin{figure} \epsscale{1.0} \plotone{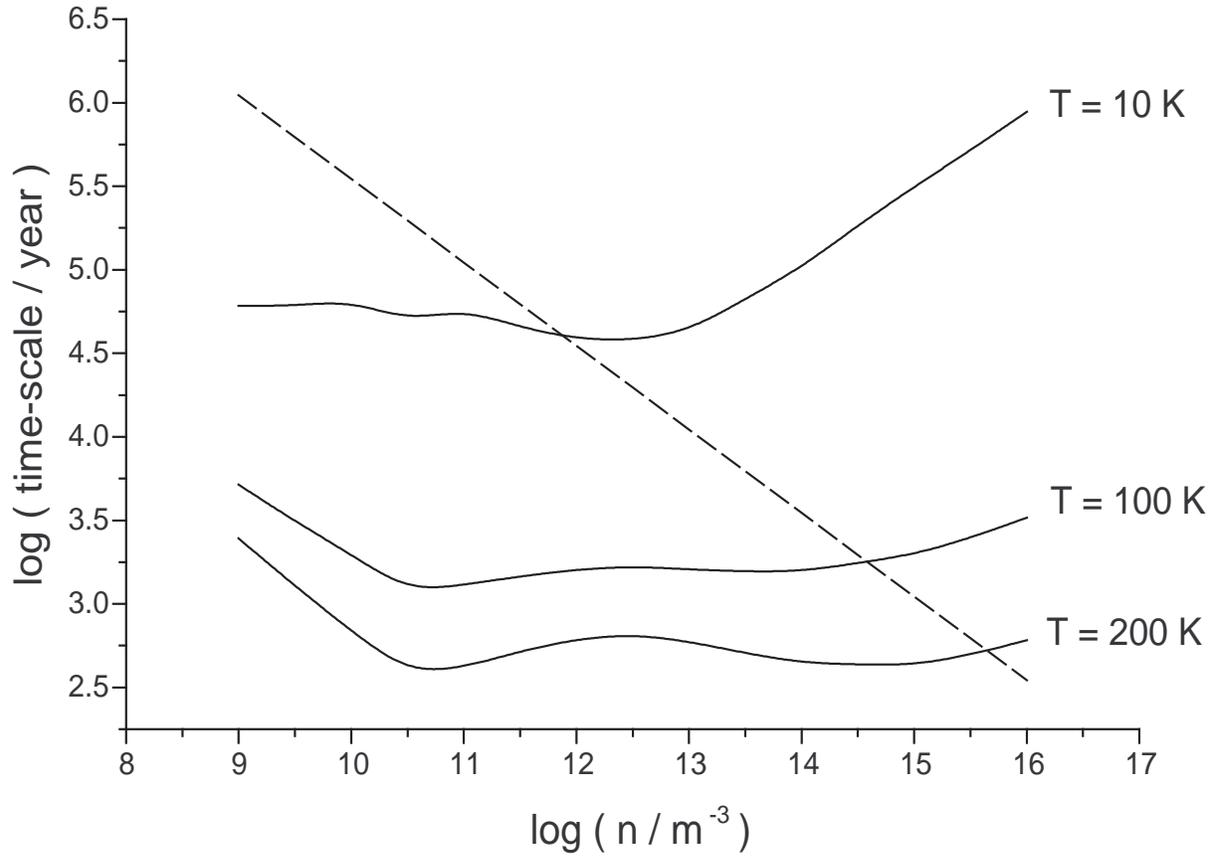}
\caption{The free-fall (dash) and cooling (solid) time-scales of a
spherical molecular cloud core for three values of
temperatures.\label{timescale}}
\end{figure}

As we know, the thermal process may be important if the dynamical
time-scale is in excess of the cooling time-scale. We consider the
contraction time-scale of a cylindrical molecular cloud as a
multiple $\eta \geq 1$ of the free-fall time-scale,
\begin{eqnarray}\label{freetime}
  \nonumber t_{ff} &=& \sqrt{\frac{3 \pi}{32 G m n}} \\
  &\approx& 3.5 \times 10^4 \left(\frac{10^{12} \mathrm{m^{-3}}}{n}\right)^{1/2}\quad \mathrm{year},
\end{eqnarray}
where the latter is written for the spherical uniform density
distribution. The cooling time-scale $t_c \equiv 3kT/2m\Lambda$ and
the free-fall time-scale (\ref{freetime}) are shown in
Fig.~\ref{timescale}. In fast contraction ($\eta \sim 1$), TI is
important in small densities (e.g., less than $10^{14} m^{-3}$ for
$T=100\mathrm{K}$), while in slow contraction ($\eta >>1$) in which
the cooling time-scale is much smaller than the contraction
time-scale, the importance of TI as a trigger mechanism is much
evident.

There are several different heating mechanisms in models of
interstellar matters. Since the ultraviolet photons are mostly
screened out in dense molecular clouds, heating by collisional
de-excitation of $H_2$ molecules after radiative excitation of Lyman
bands, photoemission from grains, radiative dissociation of $H_2$,
and by chemical reactions is not important. In addition, because of
the small neutral hydrogen abundance, heating due to ejection of
newly formed $H_2$ molecules from grain surfaces is negligible. The
heating due to cosmic rays with sufficient energies ($\sim 100
\mathrm{MeV}$) to penetrate dense clouds is commonly about
$\Gamma_{CR} = 2.5 \times 10^{-8} \mathrm{J.kg^{-1}.s^{-1}}$, with
assumption of an ionization rate per $H_2$ molecule of $2\times
10^{-17} \mathrm{s^{-1}}$ and a mean energy gains per ionization of
$19 \mathrm{eV}$ (e.g., Glassgold and Langer~1973). Following
Black~(1987) the turbulence dissipation heating rate can be
estimated as
\begin{eqnarray}\label{turb}
  \nonumber \Gamma_{TR} &\approx& \frac{(\frac{1}{2} m v_{turb}^2)
  (v_{turb}n) (\frac{1}{l})}{mn} \\
  &\approx& 1.6 \times 10^{-8} (\frac{v_{turb}}{1~\mathrm{km.s}^{-1}})^3
  (\frac{1~\mathrm{pc}}{l}) \quad
  \mathrm{J.kg}^{-1}.\mathrm{s}^{-1},
\end{eqnarray}
where $v_{turb}$ is the turbulent velocity and $l$ is the eddy
scale. With $v_{turb} \sim 1~ \mathrm{km.s}^{-1}$ and $l\sim
1~\mathrm{pc}$, we obtain $\Gamma_{TR} \sim 1.6 \times 10^{-8}
\mathrm{J.kg}^{-1}.\mathrm{s}^{-1}$, comparable to half of the
heating rate of cosmic rays. In this way, we collect the values of
these two heating rates to obtain $\sim 4.1\times 10^{-8}
\mathrm{J.kg}^{-1}.\mathrm{s}^{-1}$. Another important heating
mechanism of self-gravitating contracting cloud is the heating
produced by gravitational compression work. An estimation for this
heating can be derived directly from the rate of compression work
per particle, $pd(n^{-1})/dt$, and is given by
\begin{eqnarray}\label{hearGR}
\nonumber  \Gamma_{GR} &\approx& \frac{p}{n t_{cn}}
\\ &\approx& 3.9\times 10^{-8} (\frac{1}{\eta}) \left (\frac{T}{10
\mathrm{K}}\right ) \left ( \frac{n}{10^{12}
\mathrm{m^{-3}}} \right )^{1/2} \quad \mathrm{J.kg^{-1}.s^{-1}},
\end{eqnarray}
where we have taken $dn/dt \approx n/t_{cn}$ appropriate for a
uniform contracting spherical cloud with contraction time-scale
$t_{cn} = \eta t_{ff}$ where $\eta >>1$ is for slow contraction, and
$\eta \sim 1$ is for fast one (i.e., free-fall case).

\begin{figure} \epsscale{1.0} \plotone{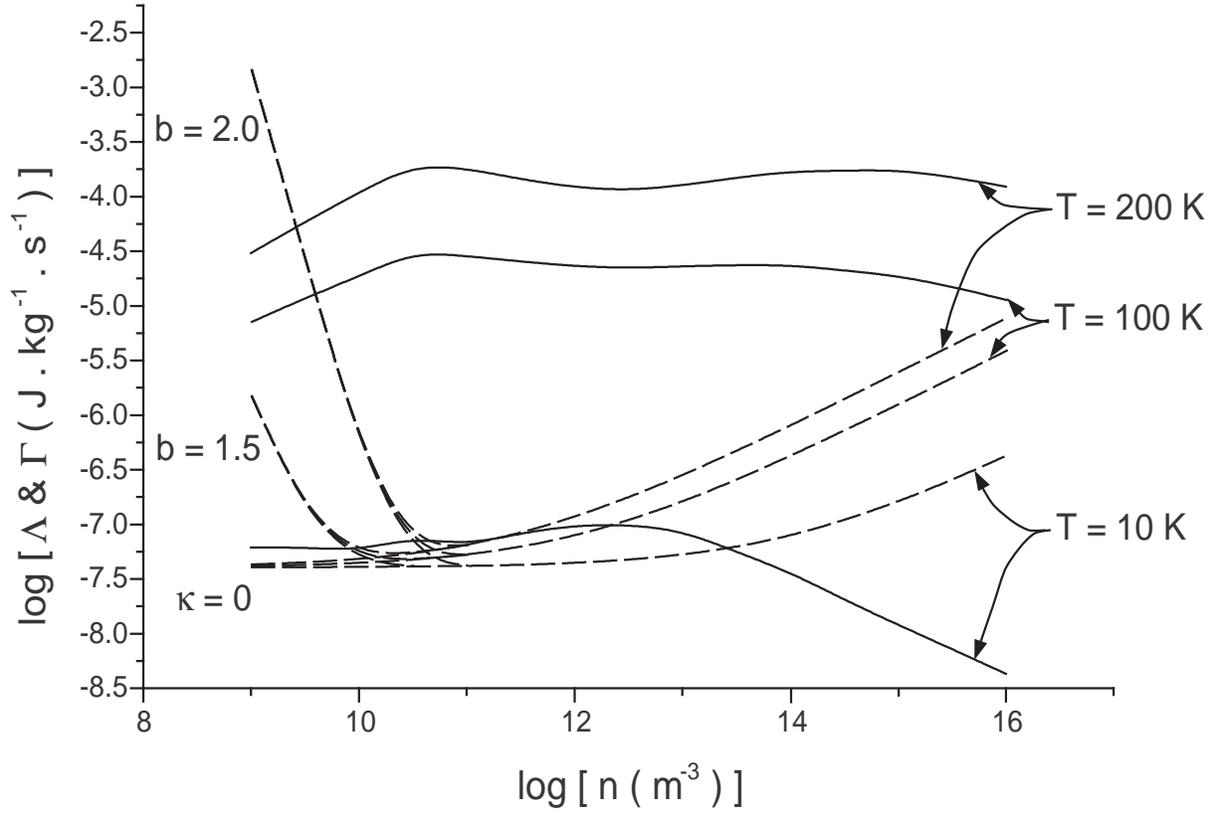}
\caption{The total cooling (solid) and heating (dash) rates by
choosing $\eta=10$ and $\kappa=0.001$ for $b=1.5$ and $2.0$. Without
heating of ambipolar diffusion (i.e., $\kappa = 0$), there would not
be any TI, while considering this heating can worthily satisfy the
Field's instability criterion (\ref{critins}) in attenuated region
of a molecular cloud.\label{coolheat}}
\end{figure}

The dissipation of magnetic energy would be considered as another
heating mechanism, if this energy is not simply radiated away by
atoms, molecules, and grains. The major field dissipation mechanism
in the dense clouds is almost certainly ambipolar diffusion, which
was examined by Scalo~(1977) for density dependency of magnetic
field in a fragmenting molecular cloud. Padoan, Zweibel and
Nordlund~(2000) presented calculations of frictional heating by
ion-neutral drift in three-dimensional simulations of turbulent,
magnetized molecular clouds. They show that average value of
ambipolar drift heating rate can be significantly larger than the
average of cosmic-ray heating rate. In addition, Nejad-Asghar~(2007)
considered a molecular slab under the assumption of
quasi-magnetohydrostatic equilibrium, concluded that ambipolar drift
heating is inversely proportional to density and its value in some
regions of the slab can be significantly larger than the average
heating rates of cosmic rays and the dissipating turbulent motions.
To gain an insight on the heating rate of ion-neutral friction, we
assume that the pressure and gravitational force on the charged
fluid component are insignificant compared to the Lorentz force
because of low ionization fraction, thus, the drift velocity $v_d$
is inversely proportional to the density and directly proportional
to the gradient of magnetic pressure (see equation~[\ref{drift}]
below). Here we choose, in a general form, $v_d \propto \kappa
\rho^{-b}$ where $\kappa \equiv \Delta (B^2/2\mu_0)/ \Delta x$ is
the change of magnetic pressure in length-scale $\Delta x$, and the
value of $b$ may be approximated near $3/2$ (this value is from the
assumption of ionization equilibrium with the ion density being
power law of the neutral density with power $1/2$).
Nejad-Asghar~(2007) adopted the value of $b$ in the range between
$0.5$ and $2.0$ to examine the isobaric TI in the regions of a
self-gravitating molecular slab. The heating due to ambipolar
diffusion is
\begin{equation}\label{heatingAD}
  \Gamma_{AD} = \frac{\textbf{f}_d.\textbf{v}_d}{\rho} =
  \gamma_{AD} \epsilon \rho^{1/2} v_d^2,
\end{equation}
where $\textbf{f}_d = \gamma_{AD} \epsilon \rho^{3/2} \textbf{v}_d$
is the drag force per unit volume exerted on the neutrals by ions,
$\gamma_{AD} \sim 3.5 \times 10^{10} \mathrm{m^3.kg^{-1}.s^{-1}}$ is
the collision drag in molecular clouds, and we used the relation
$\rho_i=\epsilon\rho_n^{1/2}$ between ion and neutral densities in
ionization equilibrium state with $\epsilon\sim 9.5\times 10^{-15}
\mathrm{kg^{1/2}.m^{-3/2}}$ (Shu~1992). Substituting the appropriate
values for typical molecular cloud cores, the ambipolar diffusion
heating rate is as follows
\begin{equation}\label{heatad}
\Gamma_{AD} \approx 4.6 \times 10^{-8} \left(\frac{\kappa}{1~
\mathrm{nT}^2.\mathrm{AU}^{-1}}\right)^2
\left(\frac{n}{10^{12}\mathrm{m^{-3}}}\right) ^{-2b+0.5}
\quad\mathrm{J.kg}^{-1}.\mathrm{s}^{-1},
\end{equation}
where the general form $v_d=\kappa \rho^{-b}/ \gamma_{AD} \epsilon$
is used according to equation (\ref{drift}) below. The total heating
and cooling rates of a magnetized slow contracting molecular cloud
core with typical value $\eta = 10$ are shown in Fig.~\ref{coolheat}
for different values of parameters $b$, $\kappa$ and temperature
$T$.

\begin{figure} \epsscale{0.75} \plotone{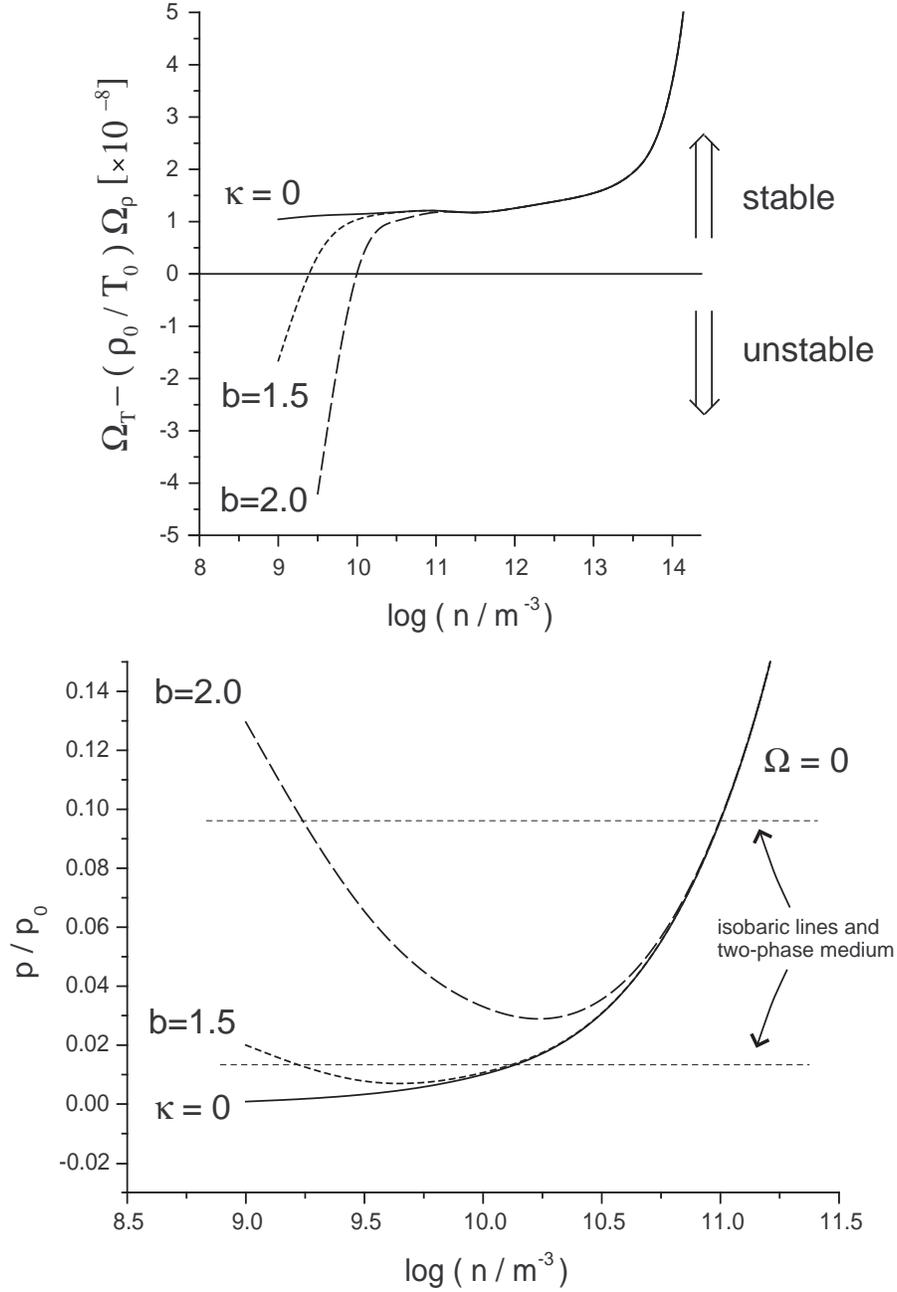}
\caption{The Field's instability criterion (top panel) and the
pressure-density diagram of thermal equilibrium (bottom panel) for
two values of parameter $b$ with $\kappa=0.001$, and without
ambipolar diffusion heating (i.e., $\kappa=0$). The pressure $p_0$
is evaluated at density $10^{12} \mathrm{m}^{-3}$ and temperature
$10 \mathrm{K}$. Occurrence of two-phase medium (constant pressure
line) in the thermally unstable region is exhibited in the bottom
panel as its instability criterion is fulfilled in the top
panel.\label{equilib}}
\end{figure}

Now, we can turn our attention to the occurrence of TI in the
molecular clouds. The standard criterion applicable to the
interstellar gas, in the isobaric instability condition, can be
written as
\begin{equation}\label{critins}
  \Omega_T - (\frac{\rho_0}{T_0})\Omega_\rho < 0,
\end{equation}
where $\Omega \equiv \Lambda - \Gamma$ is the net cooling function,
$\Omega_\rho \equiv (\partial \Omega / \partial \rho)_T$ and
$\Omega_T \equiv (\partial \Omega / \partial T)_\rho$ are evaluated
in equilibrium state (see, e.g., the review by Balbus~1995). When
the density along the net cooling curve increases so that a density
fluctuation is formed, the above criterion states it will tend to be
amplified because the mentioned fluctuation has a larger cooling
than its heating rate, thus, the surrounding gas will compress it
further. Fig.~\ref{coolheat} suggests that without heating of
ambipolar diffusion, there would not be any TI while with it, the
instability criterion (\ref{critins}) can worthily be satisfied
specially in attenuated region of a molecular cloud. Furthermore,
this scenario can also be shown by the pressure-density plane which
is an elegant method to describe the isobaric instability modes in
thermal equilibrium. In this method, an equilibrium state is
specified by the intersection of a thermal equilibrium curve and a
constant pressure line. In this way, any perturbation from
equilibrium state leads to split the gas in two phases, as the
individual gas element transfer into related phase in correspondence
with its initial sign of fluctuation (i.e., density increase or
density decrease). Pressure-density diagram of thermal equilibrium
and instability criterion are depicted in Fig.~\ref{equilib}. This
figure shows that including the heating rate due to ambipolar
diffusion can produce a two-phase medium in outer regions (low
density) of a typical molecular cloud core.

\section{Linear perturbation analysis}

For investigation the importance of TI in outer region of a
molecular cloud core, a long axisymmetric cylindrical geometry is
assumed to extend in $z$-direction with an axial magnetic field
$\textbf{B}=B_0 \hat{k}$, which directly coupled only to the charged
particles.

\subsection{Contracting cylindrical core}

Principally, the ion velocity $\textbf{v}_i$ and the neutral
velocity $\textbf{v}_n$ in the molecular clouds, should be
determined by solving separate fluid equations of these species,
include their coupling by collision processes. However, in the
time-scale considered here (see, Fig.~\ref{timescale}), two fluids
of ion and neutral are approximately coupled together with a drift
velocity given by
\begin{equation}\label{drift}
    \textbf{v}_d = \textbf{v}_i - \textbf{v}_n \approx \frac{1}{\mu_0 \gamma_{AD} \epsilon\rho^{3/2}}(\nabla\times\textbf{B})\times \textbf{B},
\end{equation}
which is obtained by assuming that the pressure and gravitational
forces on the charged fluid component are negligible compared to the
Lorentz force because of the low ionization fraction. Here, in a
good approximation we choose $\rho=\rho_n+\rho_i\approx\rho_n$. In
this way, we can use the basic equations as were given by
Shu~(1992):
\begin{equation}\label{ebmass}
\frac{\partial \rho}{\partial t} + \textbf{v} \cdot \nabla \rho +
\rho \nabla\cdot \textbf{v} =0,
\end{equation}
\begin{equation}\label{ebmoment}
\rho\frac{\partial \textbf{v}}{\partial t} + \rho (\textbf{v} \cdot
\nabla) \textbf{v} + \nabla( p + \frac{B^2}{2\mu_0}) -
(\textbf{B}\cdot\nabla)\frac{\textbf{B}}{\mu_0}-\rho \textbf{g}=0,
\end{equation}
\begin{equation}\label{ebenerg}
\frac{3}{2}\frac{\partial p}{\partial t} + \frac{3}{2} \textbf{v}
\cdot \nabla p + \frac{5}{2} p \nabla \cdot \textbf{v} +
\rho\Omega-\nabla\cdot(K\nabla T)=0,
\end{equation}
\begin{equation}\label{ebgrav}
\nabla \cdot \textbf{g}=-4\pi G \rho,
\end{equation}
\begin{equation}\label{ebstate}
p-\frac{R}{\mu}\rho T=0,
\end{equation}
where $K \approx 2.16 \times 10^{-2} T^{1/2} \mathrm{J}.
\mathrm{s}^{-1} . \mathrm{K}^{-1} . \mathrm{m}^{-1}$ is the thermal
conduction coefficient in molecular clouds (Lang~ 1986), and other
variables and parameters have their usual meanings. The energy
equation (\ref{ebenerg}) contains a source term $-\frac{5}{2}p\nabla
\cdot \textbf{v}$, which describes the work is done by expansion or
contraction of the medium. Thus, we note that the self-gravitating
heating term (\ref{hearGR}) must be excluded from the net cooling
function $\Omega$. The time evolution of magnetic field itself may
then be obtained from the approximation that it freezes only in the
plasma of ions and electrons. In this way, we obtain a nonlinear
diffusion equation as follows
\begin{equation}\label{ebmag}
\frac{\partial \textbf{B}}{\partial t}+ (\textbf{v} \cdot \nabla)
\textbf{B} + \textbf{B}(\nabla\cdot\textbf{v})-
(\textbf{B}\cdot\nabla)\textbf{v}= \nabla\times(\textbf{v}_d\times
\textbf{B} ).
\end{equation}
If the right-hand side of the magnetic induction equation
(\ref{ebmag}) is negligible, the field is well coupled to the whole
fluid.

As a basis for small-perturbation analysis, we consider an
axisymmetric cylindrical background that is contracting in a
quasi-hydrostatic balance between the self-gravity and the repulsive
forces (i.e., thermal and magnetic pressure forces). The background
quantities will be denoted with the subscript "0". We consider a
similarity solution for the bulk unperturbed fluid so that its
velocity field depends linearly on the axial distance $r$ as
$\mathbf{v}_{(\mathbf{r},t)} = \frac{ds/dt}{s} \mathbf{r}$, where
$s_{(t)}$ is a non-dimensional parameter which presents axial
contraction in the co-moving Lagrangian coordinate ($\dot{s}<0$).
The basic equations (\ref{drift})-(\ref{ebmag}) lead to
\begin{equation}\label{backgden}
\rho_{0(r,t)}= \rho_c \left(\frac{r}{r_{in}}\right)^{-\frac{2}{3}}
\left(\frac{1}{s}\right)^{\frac{4}{3}},
\end{equation}
\begin{equation}\label{backgtemp}
T_{0(r,t)}= T_c \left(\frac{1}{s}\right)^{\frac{4}{3}},
\end{equation}
\begin{equation}\label{backgmag}
B_{0(r,t)}=B_c f(\frac{r}{r_{in}})
\left(\frac{1}{s}\right)^{\frac{4}{3}},
\end{equation}
where $\rho_c$, $T_c$, and $B_c$ are respectively initial central
density, temperature, and magnetic field at the inner radius
$r_{in}$, $f(\frac{r}{r_{in}})$ is only a function of radius, and
$s_{(t)}$ follows the equation
\begin{equation}\label{st}
    s = \left[1- \frac{2}{3} (\frac{t}{t_0})\right] ^{\frac{3}{2}},
\end{equation}
where $t_0$ is a time-scale free parameter ($t \leq t_0$). The
equations (\ref{backgden})-(\ref{backgmag}) are accurate for $r >
r_{in}$, and we assume the density is homogenous for $r \leq
r_{in}$. For details, the reader is referred to the appendix. At the
end of this contraction process in which $t$ is equal to $t_0$, the
minimum value of $s_{(t)}$ takes place, thus the maximum effect of
this contraction is an approximately tenfold over density in its
profile.

\subsection{Linear perturbation}

For obtaining a linearized system of equations, we split each
variable into unperturbed and perturbed components, the latter is
indicated by subscript "1". Then, we carry out a spatial Fourier
analysis with components proportional to $\exp(ikr)$ where $k$ is
the component of perturbation wave-vector perpendicular to the
magnetic field. Simplifying the linearized equations
(\ref{drift})-(\ref{ebstate}) by repeated use of the unperturbed
background equations (\ref{backgden})-(\ref{st}), we obtain
\begin{equation}\label{linden}
    \frac{d \rho_1}{dt} + \left(ikr +1 \right) \frac{\dot{s}}{s}
    \rho_1 +\left(ikr - \frac{2}{3}\right) \frac{\rho_0}{r}  v_1 =0,
\end{equation}
\begin{equation}\label{linmom}
    \rho_0 \frac{d v_1}{dt} +\left(\frac{\ddot{s}}{s} r - g_0 + \frac{4 \pi G \rho_0 r}{ikr+1} \right) \rho_1
    + \frac{\dot{s}}{s} \rho_0 v_1 + ik p_1 + \left(\frac{ik B_0}{\mu_0} +
    \frac{1}{\mu_0} \frac{\partial B_0}{\partial r} \right) B_1 =0,
\end{equation}
\begin{equation}\label{lineng}
    \frac{3}{2} \frac{d p_1}{dt} + \left(\frac{3}{2} i k r + \frac{5}{2} \right) \frac{\dot{s}}{s}
    p_1 + \left(\frac{5}{2} ikr -1 \right) \frac{p_0}{r} v_1 + \rho_0
    \Omega_\rho \rho_1 +\left(\rho_0 \Omega_T + k^2 K_0\right) T_1=0,
\end{equation}
\begin{equation}\label{linmag}
   \frac{d B_1}{dt} + \left[\left(ikr +1\right) \frac{\dot{s}}{s} + ik v_d + \frac{1}{r}
   \frac{\partial (r v_d)}{\partial r}\right] B_1 + \left(ikB_0 +\frac{\partial B_0}{\partial r}\right) v_1
   + \frac{1}{r} \frac{\partial}{\partial r} \left( r B_0 v_{d1}\right) = 0,
\end{equation}
\begin{equation}\label{linsta}
    p_1 = \frac{R}{\mu} \rho_0 T_1 + \frac{R}{\mu} T_0 \rho_1,
\end{equation}
where
\begin{equation}\label{lindrif}
    v_{d1} = - \frac{1}{\mu_0 \gamma_{AD} \epsilon \rho_0^{3/2}}
    \left[  \left( ikB_0 + \frac{\partial B_0}{\partial r} \right) B_1 -
    \frac{3}{2} \frac{B_0}{\rho_0} \frac{\partial B_0}{\partial r} \rho_1\right].
\end{equation}
The perturbed drift velocity due to the perturbed magnetic field
changes the ambipolar diffusion heating rate. In fact, the net
cooling function $\Omega$ in equation (\ref{lineng}) must contain
also the ambipolar heating due to perturbed magnetic field. However,
because we have expressed the ambipolar heating in equation
(\ref{heatad}) by parameters, in numerical calculations for drawing
figures, we can involve this effect by the amount of parameters.
Thus, for simplicity, we neglect the explicit calculations of the
changes of ambipolar heating due to perturbed drift velocity.

Since the equilibrium is time dependent, the normal modes of the
system are time dependent too, thus, we must apply some
approximation techniques namely WKB approximation (e.g., Bora and
Baruah~2008) to gain valuable insight into the nature of problem;
this analysis may be a challenging task in a subsequent research.
Here, we consider the isobaric TI, which is more realistic phenomena
in the interstellar gases. Gathering the equations (\ref{linsta})
and (\ref{lineng}) with equation (\ref{linden}), in isobaric case
($p_1=0$), leads to an exponential growth as follows
\begin{equation}\label{expden}
   \rho_1 = \rho_{1(t=0)} \exp{\left[ - \int_0^t \omega(r,t') dt'
   \right]},
\end{equation}
where $\omega(r,t')$ is a complex function with real and imaginary
components given by
\begin{equation}\label{realomg}
    \Re [\omega(r,t')]= \frac{\dot{s}}{s} + \frac{1+\frac{3}{2}
    \frac{5}{2} k^2 r^2} {1+ (\frac{5}{2})^2 k^2 r^2} \left( k_T - k_\rho
    + \frac{k^2}{k_K}\right)c_0,
\end{equation}
and
\begin{equation}\label{imagomg}
    \Im [\omega(r,t')]= \frac{\dot{s}}{s} kr + \frac{k r} {1+ (\frac{5}{2})^2 k^2 r^2} \left( k_T - k_\rho
    + \frac{k^2}{k_K}\right)c_0,
\end{equation}
respectively, where the notation of the pioneer work of Field~(1965)
is used as follows
\begin{equation}\label{notak}
    k_T \equiv \frac{2}{3} \frac{\mu}{R} \frac{\Omega_T}{c_0},\quad
    k_\rho \equiv \frac{2}{3} \frac{\mu}{R} \frac{\rho_0\Omega_\rho}{T_0 c_0},\quad
    k_K \equiv \frac{3}{2} \frac{R}{\mu} \frac{\rho_0 c_0}{K_0},
\end{equation}
where $c_0$ is the speed of sound in the unperturbed medium. The
Field's isobaric instability criterion ($k_T < k_\rho - k^2/k_K$)
can be clearly revived via the equation (\ref{realomg}) by
considering a stationary cloud ($\dot{s}=0$).

Manipulating the equation~(\ref{realomg}) leads to the isobaric TI
criterion in a contracting axisymmetric cylindrical cloud core as
follows
\begin{equation}\label{inscriter}
    I_T(r,t)- I_\rho (r,t) + k^2 r^2 I_K (r,t) + \frac {1+ (\frac{5}{2})^2 k^2 r^2}{1+\frac{3}{2}
    \frac{5}{2} k^2 r^2} I_c(t) <0,
\end{equation}
where
\begin{eqnarray}\label{Iha}
  \nonumber  I_T(r,t) &\equiv& \frac{2}{3} \frac{\mu}{R} \int_0^{t/t_0} \Omega_T \quad d (t'/t_0), \\
  \nonumber  I_\rho(r,t) &\equiv& \frac{2}{3} \frac{\mu}{R} \int_0^{t/t_0} \frac{\rho_0\Omega_\rho} {T_0} \quad d(t'/t_0), \\
  \nonumber  I_K(r,t) &\equiv& \frac{2}{3} \frac{\mu}{R} \int_0^{t/t_0} \frac{K_0}{r^2\rho_0}\quad d(t'/t_0)\quad=\quad
    \frac{2}{3} \frac{\mu}{R} \frac{K_c t_0}{\rho_c r_{in}^2} \left( \frac{r}{r_{in}}
    \right)^{-\frac{4}{3}}
    \frac{t}{t_0} (1-\frac{1}{3} \frac{t}{t_0}), \\
  I_c(t) &\equiv&  \int_0^{t/t_0} \frac{\dot{s}}{s} \quad d(t'/t_0) \quad= \quad \frac{3}{2} \ln (1- \frac{2}{3} \frac{t}{t_0}),
\end{eqnarray}
where the $I_K$ and $I_c$ are explicitly evaluated, using the
background equations (\ref{backgden})-(\ref{st}), and $K_c \equiv
2.16 \times 10^{-2} T_c^{1/2}\mathrm{J}. \mathrm{s}^{-1} .
\mathrm{K}^{-1} . \mathrm{m}^{-1}$ is the thermal conduction
coefficient in the axis of the cylinder.

There are three notable remarks for $I_K$ and $I_c$ in the equation
(\ref{inscriter}): (\textrm{I})~Since $I_K$ has a positive value and
$I_c$ has a negative value, their effects on instability criterion
are reciprocal. Several studies have demonstrated that thermal
conduction can stabilize and even erase a density perturbation
(e.g., Burkert and Lin 2000). This implies that in the opposite of
thermal conduction, the contraction can fortify TI because the
energy is transported inward by the contraction process.
(\textrm{II})~During the growth of density perturbations in isobaric
regime, the resulting temperature gradient induces the conductive
heating in fluctuations. The temperature gradient is greater at the
distant region of the axis because in isobaric perturbation ($T_1= -
\frac{T_0} {\rho_0} \rho_1$), $\rho_0$ is a decreasing function of
the axial radius $r$. Thus, the larger temperature gradient at outer
regions leads to amplify the $I_K$ for more suppression of
perturbations. (\textrm{III})~Consideration of time dependencies of
$I_K$ and $I_c$ shows that increasing of $|I_c|$ is faster than
$I_K$ for $t > 0.5 t_0$. Thus, in general the formation of LMCs via
isobaric TI may be suppressed by the thermal conduction process as
long as it overcomes on the contraction effect.

\begin{figure} \epsscale{0.8} \plotone{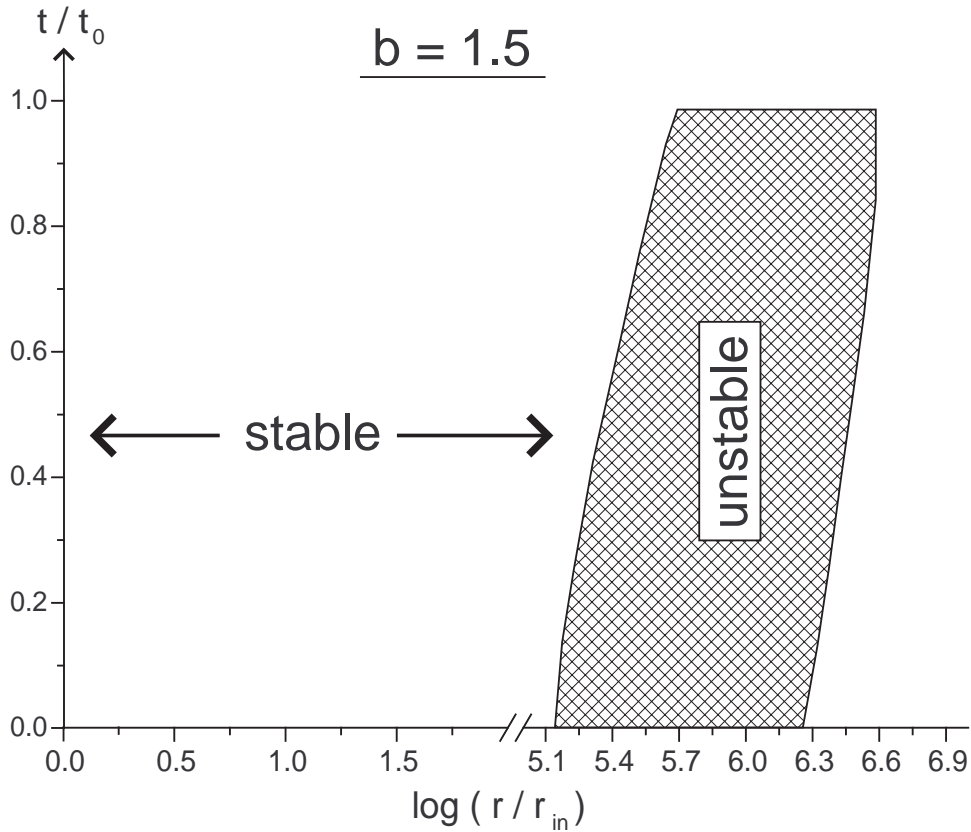}
\caption{The thermally unstable region (shade) of contracting
axisymmetric molecular cloud core including the ambipolar diffusion
with the parameter $b=1.5$, central density $n_c=4.6 \times 10^{13}
m^{-3}$, $r_{in}=0.002au$, and $t_0=1 \mathrm{Myr}$. The $k^2r^2$ is
assumed to be small so that the effect of thermal conduction in
instability criterion (\ref{inscriter}) is
ignorable.\label{inspoints}}
\end{figure}

Scrutiny in the isobaric TI criterion (\ref{inscriter}) requires to
apply some numerical values from the typical molecular cloud cores.
Here, we consider a typical core with a dimension $r_{out}=0.1
\mathrm{pc}$ and central density $n_c=4.6 \times 10^{13}
\mathrm{m^{-3}}$, which its total mass is approximately close to the
one solar mass. The central uniform region is assumed to spread out
to $r_{in}=0.002 \mathrm{AU}$. Thus, the density in the envelope
will be $n_{out}=10^9 \mathrm{m^{-3}}$, which is obtained from
initial density-profile (\ref{backgden}). Choosing $T_c=15
\mathrm{K}$ and $b=1.5$, provides a suitable situation for happening
the TI as depicted in Fig.~\ref{equilib}. The thermally unstable
region of a contracting axisymmetric cylindrical core, according to
criterion (\ref{inscriter}), is shown in Fig.~\ref{inspoints} for
$t=0$ to $t=t_0$, where it was assumed that $t_0=1 \mathrm{Myr}$. In
this figure, the effect of thermal conduction that is true in the
limit of very long wavelength ($k^2r^2 << 1$)  is ignored. We can
see from the shade region of Fig.~{\ref{inspoints}} when the
thermally unstable region of the cloud shifts to the outer parts. If
we consider an inner marginal part of the core which is thermally
unstable (for example $r =10^{5.2} r_{in}$), it may be converted to
a LMC via TI process. But, by the time, this part will be stable as
is shown in Fig.~{\ref{inspoints}}, thus, this preformed LMC will be
self-perpetuated.

\begin{figure} \epsscale{0.8} \plotone{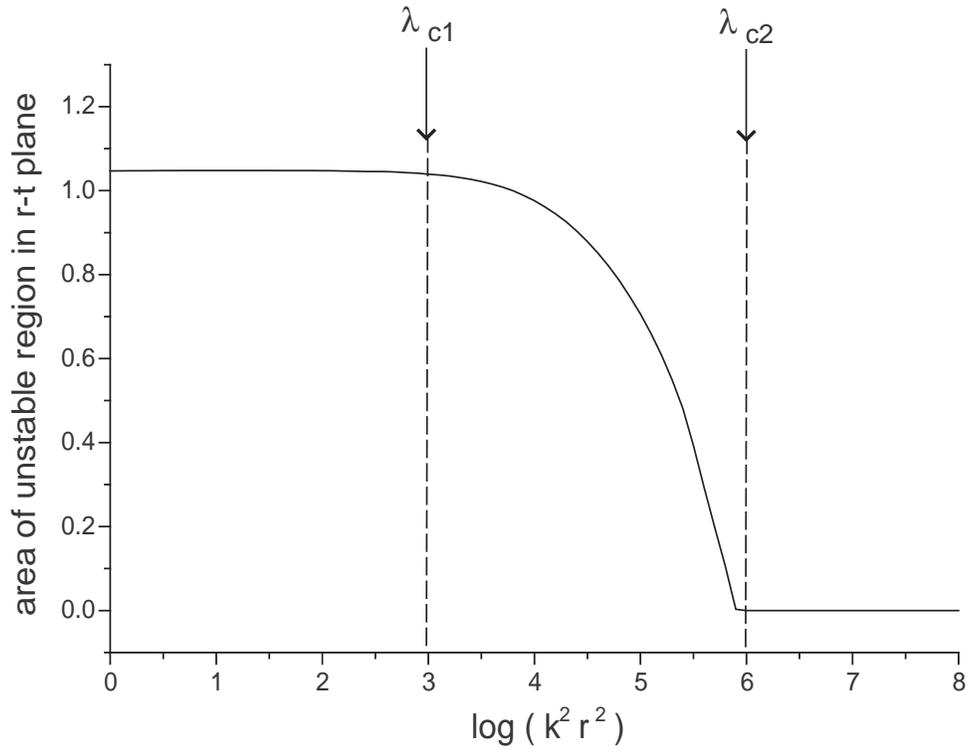}
\caption{Area of the unstable region in the $r-t$~plane of
Fig.~\ref{inspoints} versus $k^2r^2$. The thermal conduction can
entirely suppress the TI for a perturbation with wavelength less
than $\lambda_{c2}$, while its effect is completely ignorable at
wavelengths greater than $\lambda_{c1}$.\label{k2r2}}
\end{figure}

As we know, the medium may be stabilized by allowing the thermal
conduction, and its significance depends on the perturbation
wavelength which is known as Field's length (Field~1965). For this
purpose, the area of unstable region in the $r-t$~plane of
Fig.~\ref{inspoints} is depicted in Fig.~\ref{k2r2} for various
values of $k^2r^2$ (i.e., different perturbation wavelengths).
According to this figure, we can realize two critical wavelengths
that are $k_{c1}r \sim 10^{1.5}$ and $k_{c2}r \sim 10^{3}$. If we
consider the maximum value of radius as $r \sim 10^{6.5}r_{in}$, the
first critical wavelength will be $\lambda_{c1} \sim 10^5 r_{in}$.
Perturbations with wavelengths greater than the mentioned critical
value, in outer region of the molecular cores are thermally unstable
and may be considered as suitable places for the formation of LMCs.
On the other hand, using the minimum value of the radius of unstable
region, $r \sim 10^{5}r_{in}$, the second critical wavelength will
be $\lambda_{c2} \sim 10^2 r_{in}$. Perturbations with wavelengths
between $\lambda_{c1}$ and $\lambda_{c2}$ are partially affected by
thermal conduction while perturbations with wavelengths less than
$\lambda_{c2}$ will entirely be suppressed. For $k_{c2}$ we used the
maximum allowed $r$ in the unstable region and for $k_{c1}$ its
minimum was used. These choices are because of obtaining the minimum
value of wavelength for $\lambda_{c2}$ and the maximum value of it
for $\lambda_{c1}$.

It is interesting to obtain the growth time-scale of the TI as
follows
\begin{equation}\label{tigrowth}
   t_{growth} \approx \frac{\int_0^t\Re [\omega(r,t')] dt'}{t} =
   \frac{c_0}{t/t_0} \left[ I_T - I_\rho +I_c \right]
\end{equation}
where is written for the limiting case $k^2r^2 << 1$. The result is
approximately about $0.2 \mathrm{Myr}$ that is much smaller than
Jeans and magnetic instability time-scales (e.g., Fiege and
Pudritz~2000, Basu, Ciolek and Wurster~2009), as we also expected
from the Fig.~\ref{timescale}. Thus, the TI can justify the
formation of LMCs from all perturbations greater than $\lambda_{c2}$
in outer region (low density) of the molecular cloud cores.

\section{Summary and conclusions}

In this paper, we attempted to represent importance of TI as a
trigger process for configuration the hierarchical structure of
molecular clouds, specially formation of the observed LMCs in the
cloud cores. The cores are usually well thought-out as cool
molecular gases in which the temperature and density are not
completely uniform. Here, we considered the cooling rate of the
molecular cloud (Fig.~\ref{fitlambda}) and various mechanisms of
heating rates. Investigation of the net cooling rate in the
molecular clouds  shows that the consideration of the heating due to
ion-neutral drift can lead to the thermally unstable gas
(Figs.~\ref{coolheat} and \ref{equilib}). As we know, the
significance of a physical process might be regarded by its
time-scale. For this purpose, we compared the cooling time-scale and
the contracting time-scale of the core, and we concluded that in
fast contraction (free-fall), the TI is important in small
densities, while in slow contraction, the cooling time-scale is even
much shorter than the contraction time-scale (Fig.~\ref{timescale}).
Thus, the TI can be considered as a trigger mechanism to imprint the
growth of inhomogeneities within the density and temperature of the
molecular cloud cores.

For investigation the importance of the TI and finding where it will
take place, we applied the linear perturbation analysis on a
contracting axisymmetric cylindrical cloud core. Here, we turned our
attention to the isobaric TI. The obtained instability criterion
(\ref{inscriter}) demonstrates that in opposite to the thermal
conduction, which can suppress the linear density perturbations in
the medium, the contraction can fortify it because the energy is
transported inward. Furthermore, the large gradient of temperature
at outer regions of the core leads to more suppression of
perturbations therein via the thermal conduction process. Another
remark is that the formation of condensations via isobaric TI may in
general be suppressed by the thermal conduction process as long as
it overcomes on the contraction effect. Applying the numerical
values of a typical molecular cloud core gave us some insight about
the places where instability may be arisen. Fig.~\ref{inspoints}
depicts the thermally unstable regions of a contracting axisymmetric
cylindrical core in the limitation of very long wavelength (i.e.,
the effect of thermal conduction is ignorable). According to this
figure, by the time, unstable region of the cloud shifts to the
outer region, the preformed LMCs can be self-perpetuated.

The medium would be stabilized by consideration of thermal
conduction which its significance depends on perturbation wavelength
(Fig.~\ref{k2r2}). The thermal conduction can entirely suppress the
instability of perturbations with wavelengths less than
$\lambda_{c2} \sim 0.2 \mathrm{AU}$, while its effect is completely
ignorable at wavelengths greater than $\lambda_{c1} \sim 200
\mathrm{AU}$. Thus, LMCs more massive than $m_{H_2} \overline{n}
\lambda_{c1}^3 \sim 10^{25} \mathrm{kg}$ can grow via TI without
being undergone the thermal conduction, and those smaller than
$m_{H_2} \overline{n} \lambda_{c2}^3 \sim10^{16} \mathrm{kg}$ are
completely disrupted by the thermal conduction process
($\overline{n}$ is chosen approximately $10^{10} \mathrm{m}^{-3}$
where TI occurs). Perturbations with having wavelengths between
$\lambda_{c1}$ and $\lambda_{c2}$ are partially affected by the
thermal conduction. The relative mass which is contained within the
thermally unstable region of the cylindrical core is approximately
$\frac{\pi r_{max}^2 - \pi r_{min}^2}{\pi r_{max}^2} = 1-
\left(\frac{r_{min}} {r_{max}}\right)^2 \approx 0.99$, where
$r_{min} \approx 10^{5.15} r_{in}$ and $r_{max}\approx 10^{6.25}
r_{in}$ are chosen from Fig.~\ref{inspoints}. Thus, since
approximately $99\%$ of the core mass is contained within the
thermally unstable region, its mass is sufficient to explain the
formation of LMCs. Although, the TI may justify the formation of
LMCs more massive than $10^{-5} M_\odot$, however, the puzzle is
highly incomplete. We might investigate the behavior and evolution
of these LMCs because they move and accumulate ubiquitous dust and
other molecules. Furthermore, they may collide and merge to form the
larger LMCs. Finally, they may be converted to collapsing
proto-stellar entities, brown dwarfs, or proto-planets.

\section*{Acknowledgments}
I appreciate the careful reading and suggested improvements by Sven
Van Loo, the reviewer. This work has been supported by Research
Institute for Astronomy and Astrophysics of Maragha (RIAAM).
\appendix

\section{Background evolution}
In order to solve the set of equations (\ref{ebmass})-(\ref{ebmag}),
we look for a similarity contraction using the co-moving system:
$\textbf{r} = s_{(t)} \textbf{x}$, where $\textbf{x}$ is the
Lagrangian coordinate and $s_{(t)}$ is the contraction parameter. In
this way, the velocity field is given by
$\textbf{v}_{(\textbf{r},t)} = \frac{ds/dt}{s} \textbf{r}$. From the
equation of continuity (\ref{ebmass}), we have
\begin{equation}\label{apmass}
    \frac{\partial \rho_0}{\partial t} + \frac{\dot{s}}{s} r \frac{\partial \rho_0}{\partial r} + 2 \frac{\dot{s}}{s} \rho_0 = 0.
\end{equation}
Using the separation of variables to solve the equation (\ref{apmass}), we obtain
\begin{equation}\label{apden}
    \rho_{0(r,t)}= \rho_c \left(\frac{r}{r_{in}}\right)^{a-2} \left[\frac{1}{s_{(t)}}\right]^{a},
\end{equation}
where $\rho_c$ is the initial central density and the constraint
$0<a<2$ fulfills the decreasing of density versus the cylindrical
radius and its increasing with time. The equation (\ref{apden}) is
accurate for $r
> r_{in}$, and we assume the density is homogenous (i.e. $a=2$) for
$r \leq r_{in}$.

Substituting the equation (\ref{apden}) into the poisson equation
\begin{equation}\label{appois}
    \frac{1}{r} \frac{\partial}{\partial r} (rg_0) = -4 \pi G \rho_0,
\end{equation}
we obtain, for $r > r_{in}$,
\begin{equation}\label{apgrav}
    g_{0(r,t)} = - g_c \left(\frac{r}{r_{in}}\right)^{a-1} \left[\frac{1}{s_{(t)}}\right]^{a},
\end{equation}
where $g_c\equiv 4 \pi G \rho_c r_{in} /a$ is the initial
gravitational acceleration at $r_{in}$ for a long homogenous
cylinder, and we set $a=2$ for $r \leq r_{in}$, as before. We assume
that the temperature of the background is homogenous and it depends
only on the time so that $p_{(r,t)}=\frac{R}{\mu} \rho_{(r,t)} T
_{(t)}$. From the energy equation (\ref{ebenerg}) with the
On-The-Spot equilibrium in which the energy-gain is locally equal to
the energy-loss at that spot (i.e., $\Omega_0 \approx 0$), we have
\begin{equation}\label{apenergy}
    \frac{3}{2}\frac{1}{T_0} \frac{dT_0}{dt} + \frac{3}{2} \frac{1}{\rho_0} \frac{\partial \rho_0}{\partial t} + \frac{3}{2} \frac{\dot{s}}{s} \frac{r}{\rho_0} \frac{\partial \rho_0}{\partial r} + 5 \frac{\dot{s}}{s} =0,
\end{equation}
which gives
\begin{equation}\label{aptemp}
    T_{0(r,t)} = T_c \left[\frac{1}{s_{(t)}}\right]^{\frac{4}{3}},
\end{equation}
where $T_c$ is the initial homogenous temperature.

Separating the time-variable from the equations of momentum
(\ref{ebmoment}) and magnetic induction (\ref{ebmag}) by desiring
the magnetic field as
\begin{equation}\label{apmag}
    B_{0(r,t)}=B_c f(\frac{r}{r_{in}}) \left[\frac{1}{s_{(t)}}\right]^{b},
\end{equation}
where $B_c$ is the initial central magnetic field strength and
$f(\frac{r}{r_{in}})$ is a function of radius, we obtain the
constraint $a = b = \frac{4}{3}$ and the contraction parameter as
\begin{equation}\label{apst}
    s_{(\frac{t}{t_0})} =  \left[1- \frac{2}{3} (\frac{t}{t_0})\right] ^{\frac{3}{2}},
\end{equation}
where $t_0$ is a time-scale free parameter ($0 \leq t \leq t_0$). It
is worth notice that the equations (\ref{apden}), (\ref{aptemp}),
and (\ref{apmag}) show explicitly that the ratio of magnetic
pressure to gas pressure will be constant in duration of
contraction. Clearly, this special case is emerged from our linear
similarity assumption of the velocity field in the contracting
cylinder. The separated $r$-variable of the momentum equation
(\ref{ebmoment}) reduces to
\begin{equation}\label{apfr}
f_(\frac{r}{r_{in}}) = \left\{ f_0^2 - \frac{2 \mu_0}{B_c^2}
\frac{3\rho_c r_{in}^2}{4 t_0^2} \left(\frac{r}{r_{in}}\right) ^
{\frac{2}{3}}\left[ \frac{1}{3} \left( \frac{r}{r_{in}}
\right)^{\frac{2}{3}} +\frac{4 R T_c t_0^2 }{3\mu
r_{in}^2}\left(\frac{r}{r_{in}}\right) ^ {-\frac{4}{3}}  + \frac{2
g_c t_0^2}{r_{in}} \right] \right\}^{\frac{1}{2}}
\end{equation}
where the constant $f_0$ is chosen so that the $f(\frac{r}{r_{in}})$
be a real function. The equation (\ref{apfr}) may be inserted into
the separated $r$-variable of the magnetic induction equation
(\ref{ebmag}) to obtain $\textbf{v}_d$ via integral
\begin{equation}\label{apvd}
    \textbf{v}_d = - \hat{r} \left [ \frac{1}{s_{(t)}} \right
    ]^{\frac{2}{3}} \frac{1}{t_0 r f} \int (\frac{2}{3}f + r \frac{df}{dr}) r dr,
\end{equation}
instead the approximate relation (\ref{drift}).


\clearpage

\clearpage

\clearpage

\clearpage

\clearpage

\clearpage

\end{document}